\documentclass[aps,prb,twocolumn,showpacs, superscriptaddress]{revtex4-1}

\usepackage{amsmath,amssymb}
\usepackage{color}
\usepackage{nicefrac}
\usepackage{xspace}
\usepackage{graphicx}
\usepackage{dcolumn}
\usepackage{bm}

\newcommand{\lfmaso}{LaFe$_{1-x}$Mn$_{x}$AsO$_{0.89}$F$_{0.11}$\xspace}
\newcommand{\lfaof}{LaFeAsO$_{0.89}$F$_{0.11}$\xspace}
\newcommand{\laopt}{LaFeAsO$_{0.89}$F$_{0.11}$}
\newcommand{\lafdoped}{LaFeAsO$_{1-x}$F$_{x}$}
\newcommand{\smfaof}{SmFeAsO$_{1-x}$F$_{x}$}
\newcommand{\bfamn}{Ba(Fe$_{1-x}$Mn$_{x}$As)$_2$}
\newcommand{\bkfa}{Ba$_{0.5}$K$_{0.5}$(Fe$_{1-x}$Mn$_{x}$As)$_2$}

\newcommand{\slrr}{$T_1^{-1}$}
\newcommand{\slrrt}{$(T_1T)^{-1}$}
\newcommand{\as}{$^{75}$As}

\newcommand{\F}{Fig.~}

\begin{document}

\bibliographystyle{apsrev4-1}

\preprint{APS/123-QED}

\title{The poisoning effect of Mn in LaFe$_{1-x}$Mn$_x$AsO$_{0.89}$F$_{0.11}$: unveiling a quantum critical
point in the phase diagram of iron-based superconductors}
\author{F.~Hammerath}
\email[Corresponding author: ]{franziska.hammerath@unipv.it}
\altaffiliation[Present address: ]
{IFW-Dresden, Institute for Solid State Research, PF
270116, 01171 Dresden, Germany.}
\affiliation{Dipartimento di Fisica and Unit\`a CNISM di Pavia, I-27100 Pavia, Italy}
\author{P.~Bonf\`a}
\affiliation{Dipartimento di Fisica and Unit\`a CNISM di Parma, I-43124 Parma, Italy}
\author{S.~Sanna}
\affiliation{Dipartimento di Fisica and Unit\`a CNISM di Pavia, I-27100 Pavia, Italy}
\author{G.~Prando}
\altaffiliation[Present address: ]
{IFW-Dresden, Institute for Solid State Research, PF
270116, 01171 Dresden, Germany.}
\affiliation{Dipartimento di Fisica and Unit\`a CNISM di Pavia, I-27100 Pavia, Italy}
\author{R.~{De~Renzi}}
\affiliation{Dipartimento di Fisica and Unit\`a CNISM di Parma, I-43124 Parma, Italy}
\author{Y.~Kobayashi}
\affiliation{Department of Physics, Division of Material Sciences, Nagoya University, Furo-cho, Chikusa-ku, Nagoya 464-8602, Japan}
\author{M.~Sato}
\affiliation{Department of Physics, Division of Material Sciences, Nagoya University, Furo-cho, Chikusa-ku, Nagoya 464-8602, Japan}
\author{P.~Carretta}
\affiliation{Dipartimento di Fisica and Unit\`a CNISM di Pavia,
I-27100 Pavia, Italy}

\begin{abstract}
A superconducting-to-magnetic transition is reported for \lfaof\
where a per thousand amount of Mn impurities is dispersed. By
employing local spectroscopic techniques like muon spin rotation
($\mu$SR) and nuclear quadrupole resonance (NQR)
on compounds with Mn contents ranging from $x=0.025$\,\% to $x=0.75\,\%$, we find that the electronic
properties are extremely sensitive to the Mn impurities.
In fact, a small amount of Mn as low as 0.2\,\% suppresses superconductivity completely. 
Static magnetism, involving the FeAs planes, is observed to arise for $x> 0.1$\,\% and becomes further enhanced upon increasing Mn substitution. Also a
progressive increase of low energy spin fluctuations, leading to
an enhancement of the NQR spin-lattice relaxation rate \slrr, is
observed upon Mn substitution. The analysis of \slrr\ for the
sample closest to the the crossover between superconductivity and
magnetism ($x=0.2$\,\%) points towards the presence of an
antiferromagnetic quantum critical point around that doping level.
\end{abstract}


\maketitle

\section{Introduction}
The study of the effect of impurities on a superconductor is a
well-known and versatile method to investigate the symmetry of the
order parameter and the related pairing mechanisms.\cite{Wang2013}
Accordingly, the effects of transition metal ion
substitution{\cite{SatoJPSJ2010, Satomi2010, Kawamata2011,
Kitagawa2011Zn}} or the introduction of
deficiencies\cite{FuchsPRL2008, HammerathPRB2010, Kito2008} on the
superconducting ground state of iron-based superconductors have been
intensively studied during the last years. The superconductors of
the {{\it Ln}Fe$_{1-x}$M$_x$AsO$_{1-y}$F$_y$  ({\it Ln}1111)
family, with {\it Ln} = La, Ce, Nd, Sm, ... and M = impurity
elements doped on the Fe site, are one of the example systems used
in such studies.}

The behavior of the superconducting transition temperature
$T_{c}$ in optimally {F} doped (y $\simeq 0.11$) {\it Ln}1111
superconductors have been 
investigated under a variety of
transition metal substitutions (e.g. M = Co, Ni, and
Ru).\cite{SatoJPSJ2010, Satomi2010, Kawamata2011} The initial
suppression rates, $|{\rm d}T_c/{\rm d}x|_{x\rightarrow 0}$, are
much smaller than those typically induced by non-magnetic
impurities in systems with an $s_{\pm}$ symmetry of the
superconducting order parameter. One has to consider that $T_c$ is
primarily determined by the number of conducting
electrons\cite{Kawamata2011, Sato2012} and, indeed, Co and Ni for
Fe substitution do introduce electrons in
La1111.\cite{SatoJPSJ2010, Sato2012} On the other hand, the very
small value of $|{\rm d}T_c/{\rm d}x|_{x\rightarrow 0}$ observed
for M = Ru, a substitution which does not change the carrier
density, can be considered as an evidence that the scattering by
non-magnetic impurities does not act as an efficient pair breaking
center. The Ru substitution is also observed to induce static
magnetism for $x> 10$\%,\cite{Sanna2011, Sanna2013} indicating the importance
of the change of the electronic state caused by relatively high
doping levels.
There are also many reports on the effect of Zn doping in
Fe-based superconductors\cite{Kitagawa2011Zn, Li2010, Cheng2010,
Li2012Ba122} and on the relevance of the observed electron
localization taking place at low temperature.\cite{SatoJPSJ2010}

At variance with the cases of M = Co and Ni, a
remarkable increase of the resistivity and a very rapid
suppression of $T_c$ were found for M = Mn in optimally F doped La1111 (y =
0.11).\cite{SatoJPSJ2010} This trend has been explained
by considering the electron localization induced by Mn. Similar
effects have been observed also in \bkfa.\cite{Cheng2010,
Li2012Ba122} It has to be remarked that in contrast to other
transition metals such as Co, Mn substitution in the undoped
(antiferromagnetic) parent compound BaFe$_2$As$_2$ just leads to
a decrease of the magnetic transition temperature $T_N$, without
inducing superconductivity.\cite{Thaler2011} Several experimental
techniques such as nuclear magnetic resonance (NMR), inelastic
neutron scattering and photoemission spectroscopy showed that no
charge doping occurs upon Mn substitution and that Mn moments tend
to localize, suggesting that the moments are acting as magnetic scattering
centers.\cite{Texier2012, Suzuki2013Ba122Mn, Urata2013,
Inosov2013}
Also in undoped (antiferromagnetic) LaFeAsO, Mn magnetic moments affect the long range magnetic order within the Fe planes, which evolves into a short range magnetic order upon adding Mn, without leading to the onset of
superconductivity.\cite{Frankovsky2013}

Here we focus on the peculiar case of Mn substituion on the Fe
site in nominally optimally doped \laopt. Among the {\it Ln}1111
family, \lafdoped\ is the system with the lowest $T_c$ at optimal
doping and, remarkably, also the lowest $T_N$ of magnetic phase
induced by Ru substitution.\cite{LuetkensNatMat2009, Sanna2013}
Furthermore, no coexistence region of magnetism and
superconductivity is found in its phase diagram upon electron
doping.\cite{LuetkensNatMat2009} The low $T_c$ and $T_N$ values
indicate weaker superconducting and magnetic ground states
suggesting that \lafdoped\ is the most promising candidate to observe a quantum critical point (QCP), an aspect, which has already been pointed out at a very early stage of the pnictide research.\cite{Mazin2008} Accordingly, slight changes in the ground state, such as those induced by the insertion of low amounts of impurities, might lead
to big effects on the ground state and on the electronic
properties. 

Indeed, we observe a drastic suppression of $T_{c}$ in a very
small substitutional range, where charge doping, if any, can be
safely neglected.
Two competing magnetic and superconducting ground states are found and studied in detail by means of nuclear quadrupole resonance (NQR) and muon spin rotation ($\mu$SR) spectroscopy,
allowing us to draw the electronic phase diagram for \lfmaso\ in the low ''doping" region. We find that superconductivity is already completely
suppressed for $x=0.2$\,\%. Short range static magnetism sets in for $x\geq0.1$\,\% and becomes more and more enhanced upon further Mn substitution. \as\ NQR spin-lattice relaxation rate measurements sense a progressive slowing down of low energy spin fluctuations with increasing Mn content. The analysis of the spin dynamics within the framework of Moriya's self consistent renormalization (SCR) theory points towards the presence of a QCP at the crossover region between superconductivity and magnetism in \lfmaso.

\section{Sample Characterization}

We studied polycrystalline samples of \lfmaso\ with Mn contents of
$x=0$\,\%, $0.025$\,\%, $0.075$\,\%, $0.1$\,\%, $0.2$\,\%,
$0.5$\,\%, and $0.75$\,\%. The sample preparation and
characterization by means of electrical resistivity, Hall
coefficient, thermoelectric power and specific heat measurements
have already been discussed in Ref.~\onlinecite{SatoJPSJ2010}. The
superconducting transition temperature $T_c$ was determined via
superconducting quantum interference device (SQUID) magnetometry.
All the samples are optimally electron doped with a nominal
fluorine content of $11$\,\%. For $x\leq 0.2$\,\% $^{19}$F-NMR
measurements have been performed in an applied magnetic field of
$\mu_0 H = 1$\,T to check the relative fluorine doping level.
Within the error bars, no variation of the intensity of the
$^{19}$F-NMR resonance line was found, confirming that the
intrinsic F content does not differ among the samples within $\pm
0.005$. This emphasizes that the effects presented in the
following clearly stem from the influence of the Mn impurities
only.

The superconducting transition temperature $T_c$ was checked additionally
by following the detuning of the NQR resonance coil. $T_c=$29, 25,
16.3 and 11.5\,K were found for $x=0$\,\%, $0.025$\,\%,
$0.075$\,\%, and  $0.1$\,\%, respectively, in nice agreement with
magnetization and TF--$\mu$SR measurements.

\section{Technical Aspects and Experimental Results}

\subsection{Nuclear Quadrupole Resonance}

\begin{figure}[t]
\begin{center}
\includegraphics[width=\columnwidth,clip]{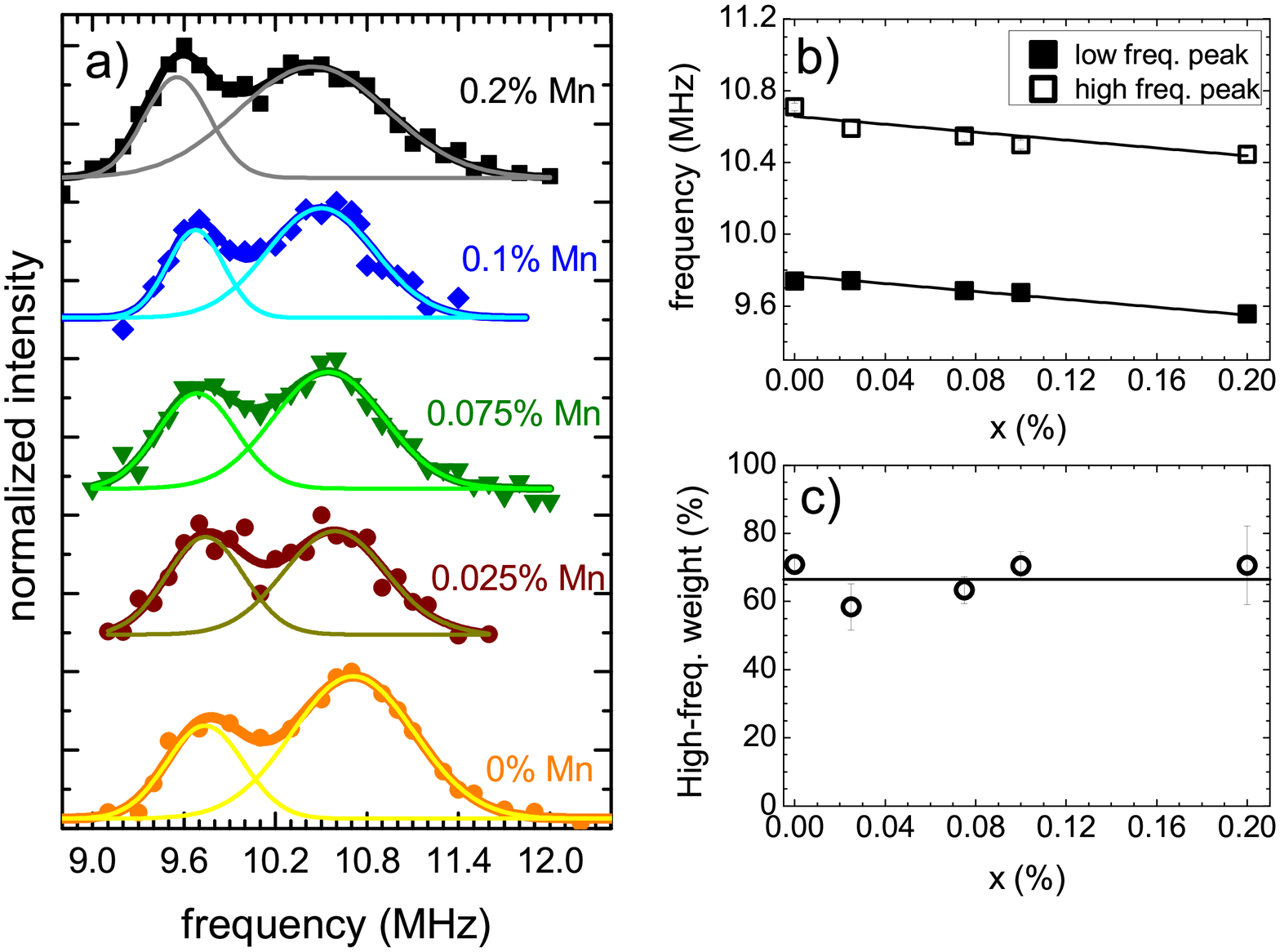} 
    \caption{(a) \as-NQR spectra of \lfmaso\ for Mn contents from $x=0$\,\% up to $x=0.2$\,\% (symbols), measured at $T=77$\,K. Solid lines are fits including two Gaussian lines for each sample. Right side: Mn-content dependent spectral peak frequencies (b) and high frequency weight (c) of the double-peaked \as-NQR spectra, deduced from the fits in (a). Filled squares in (b) denote the low frequency peak, open squares the high frequency peak. Solid lines in (b) and (c) are guides to the eyes.}\label{fig:spectra}
\end{center}
\end{figure}

\as\ nuclear quadrupole resonance (NQR) spectra allow to probe the
local charge distribution in the FeAs planes and therewith  to
evidence a possible charge doping induced by Mn. In fact, since the nuclear
quadrupole moment $Q$ of \as\ (nuclear spin $I=3/2$) interacts
with the components $V_{\alpha\beta}$ of the electric field
gradient (EFG) generated by the surrounding charge distribution, the NQR frequency turns out to be:
\begin{equation}
\nu_{NQR} = \frac{3eQV_{zz}}{2I(2I-1)h}\sqrt{1+\eta^2/3} \, ,
\end{equation}
where $V_{zz}$ and $\eta$ are the highest eigenvalue of the EFG
and its asymmetry, respectively.

For the $^{75}$As NQR measurements all samples were ground to a
fine powder to enhance radiofrequency penetration. $^{75}$As NQR
spectra were taken by integrating the full spin echo obtained
after a standard Hahn spin echo pulse sequence of the form
$\frac{\pi}{2} - \tau - \pi$ upon varying the irradiation frequency. The pulsewidth, the repetition rate
of the pulse sequences and $\tau$ were adjusted to maximize the
spin echo intensity and minimize spin-lattice and spin-spin
relaxation effects on the spectra. Fig.~\ref{fig:spectra}(a) shows
the \as-NQR spectra measured at $T=77$\,K for Mn contents from
$x=0$\,\% up to $x=0.2$\%. We observe a double-peaked \as-NQR
spectrum, very similar to what has been previously observed for
slightly underdoped \lafdoped\ samples,\cite{LangPRL2010, Oka2011,
Kobayashi2010}  where it has been assigned to two charge
environments which are coexisting at the
nanoscale.\cite{LangPRL2010} In these previous studies the shape
of \as-NQR spectrum has been found to depend strongly on the
fluorine doping level, which corresponds to effective electron
doping and affects the EFG drastically. The very similar shape of
all the spectra shown in Fig.~\ref{fig:spectra} confirms that the
fluorine content, although possibly
lower than the nominal one, does not change among the samples. 
The spectra could be well fitted with two Gaussian lines, shown
as solid lines in Fig.~\ref{fig:spectra}. The relative
weight of both Gaussians (roughly 30\,\% vs 70\,\% for the low/high frequency peak, respectively) does not change upon increasing the Mn content [see Fig.~\ref{fig:spectra}(c)]. Also the full width at half maximum (FWHM) does not change upon Mn substitution. Only for the highest measured doping level
($x=0.2$\,\%) we observe a slight broadening of the high frequency
peak, which can be possibly related to the enhanced magnetic
correlations in this compound. The slight decrease of the peak
frequencies upon increasing the Mn content [see
Fig.~\ref{fig:spectra}(b)] can be ascribed to lattice strain
associated to the presence of disorder.\cite{LangPhysicaC2010}
We do not find any evidence for a difference among the carrier numbers of these samples in the studied doping range. 
This is in agreement with previous experimental observations on Mn substituted
pnictides.\cite{Texier2012, Suzuki2013Ba122Mn, Urata2013,
Inosov2013, Ding2013} 

\begin{figure}[t]
\begin{center}
\includegraphics[width=\columnwidth,clip]{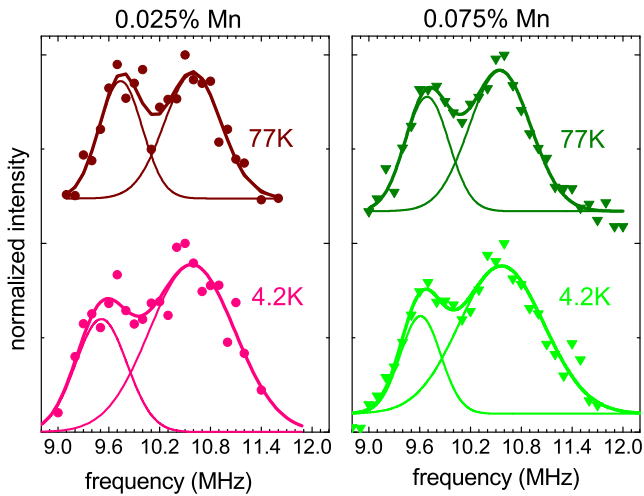} 
   \caption{Temperature dependent \as-NQR spectra of \lfmaso\ with $x=0.025$\,\% and $x=0.075$\,\%. Solid lines denote fits with two Gaussians for each spectrum.}\label{fig:spectra_Tdep}
\end{center}
\end{figure}

The temperature dependence of the shape of the \as-NQR spectra was checked for some representative samples ($x=0.025$\,\% and $x=0.075$\,\%). The results are plotted in Fig.~\ref{fig:spectra_Tdep}. While the peak frequencies and the FWHM of the low frequency peak do not change upon cooling, the FWHM of the high frequency peak increases slightly. Thus, the high frequency peak seems to be more sensitive to the growing magnetic correlations in these compounds (for the discussion of the magnetic correlations see the following discussions of the \as-NQR \slrrt\ and of the $\mu$SR results).

\begin{figure}[t]
\begin{center}
\includegraphics[width=\columnwidth,clip]{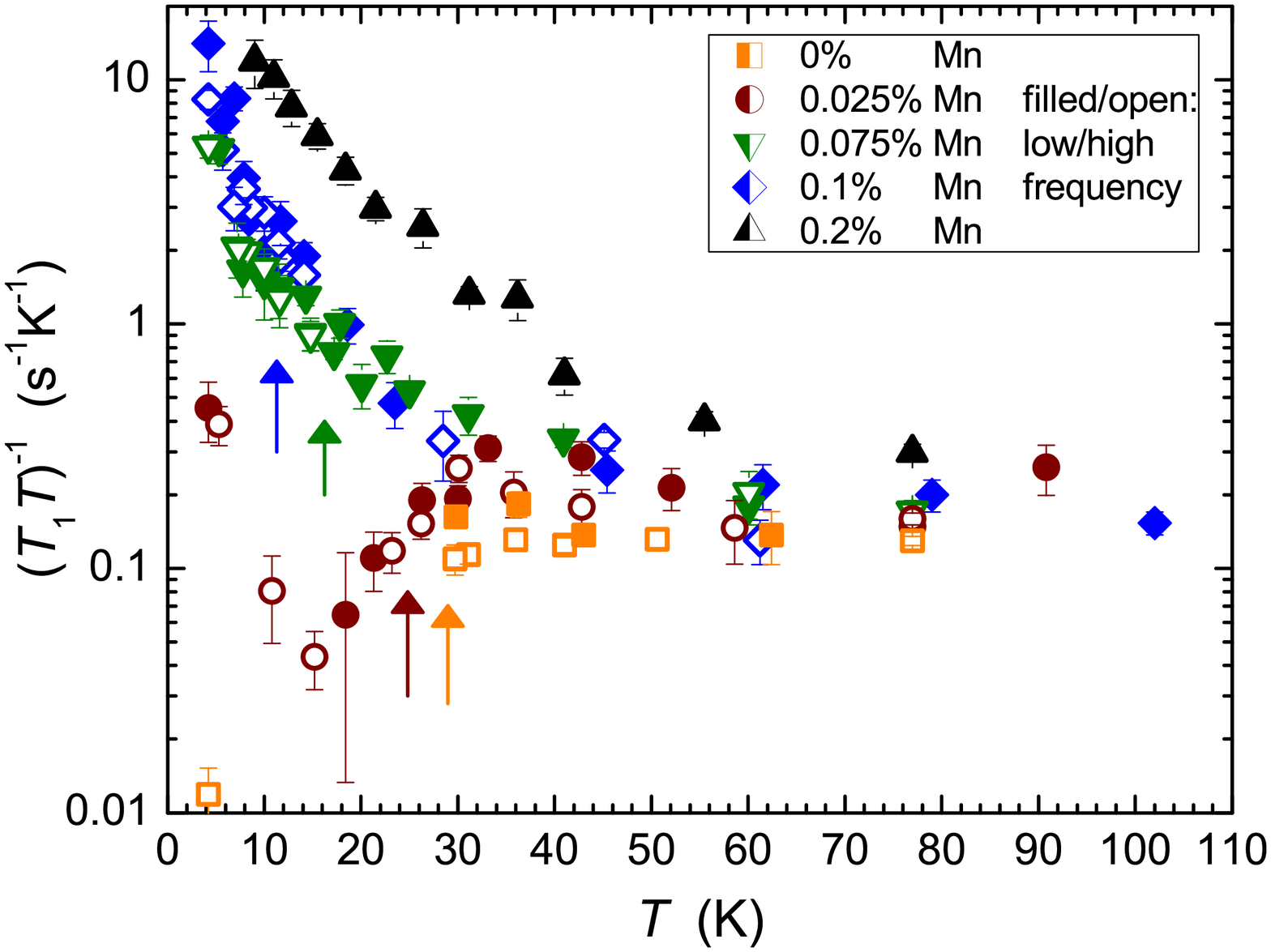} 
    \caption{(a) \as-NQR spin-lattice relaxation rate divided by temperature \slrrt\ for \lfmaso\ with $x=0$\,\% up to $x=0.2$\,\%. \slrrt\ was measured at the low and high frequency peak of the double-peaked \as-NQR spectra (filled and open symbols, respectively). For $x=0.2$\,\% only \slrrt\ of the low frequency peak was measured. The arrows denote the superconducting transition temperatures for $x=0$\,\% (orange, $T_c = 29$\,K) up to $x=0.1$\,\% (blue, $T_c = 11.5$\,K).
}\label{fig:T1T}
\end{center}
\end{figure}

The $^{75}$As NQR spin-lattice relaxation rate \slrr\ was measured with an
inversion recovery pulse sequence and the recovery of the nuclear
magnetization $M_z(\tau)$ was fitted to:
\begin{equation}
M_z(\tau) = M_0[1-fe^{-(3\tau/T_1)^\beta}] \, , \label{eq:Mz}
\end{equation}
where $M_0$ is the saturation magnetization in thermal
equilibrium, $f$ close to 2 accounts for incomplete inversion and
$\beta$ is a stretched exponent which indicates a distribution of
\slrr. In Fig.~\ref{fig:T1T} the \as\ NQR spin-lattice relaxation
rate divided by temperature \slrrt, measured for samples with
$x=0$\,\% up to $x=0.2$\,\%, is shown. At high temperatures, the
recovery of the nuclear magnetization is single exponential
[$\beta = 1$, see Eq.~\eqref{eq:Mz}] until the system reaches the
region where magnetic fluctuations start to slow down. Below
around 30 - 50\,K, depending on the doping level, $0.4 \leq \beta
\leq 0.8$ had to be used to fit the recovery.

\begin{figure*}[t]
    \includegraphics[width=0.9\columnwidth, clip]{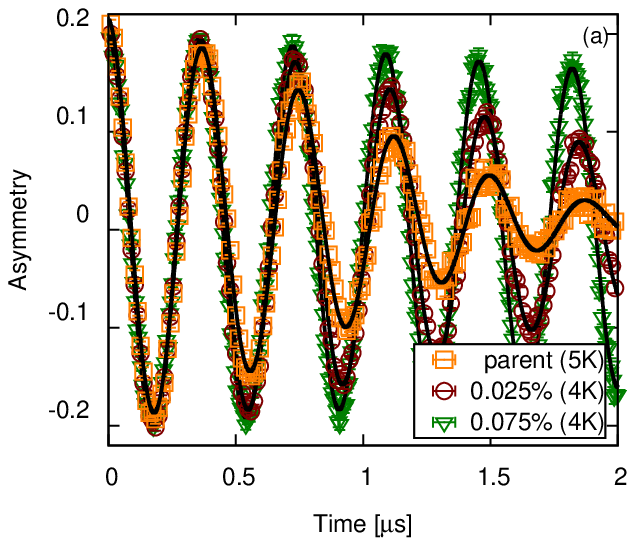} 
    \includegraphics[width=1\columnwidth, clip]{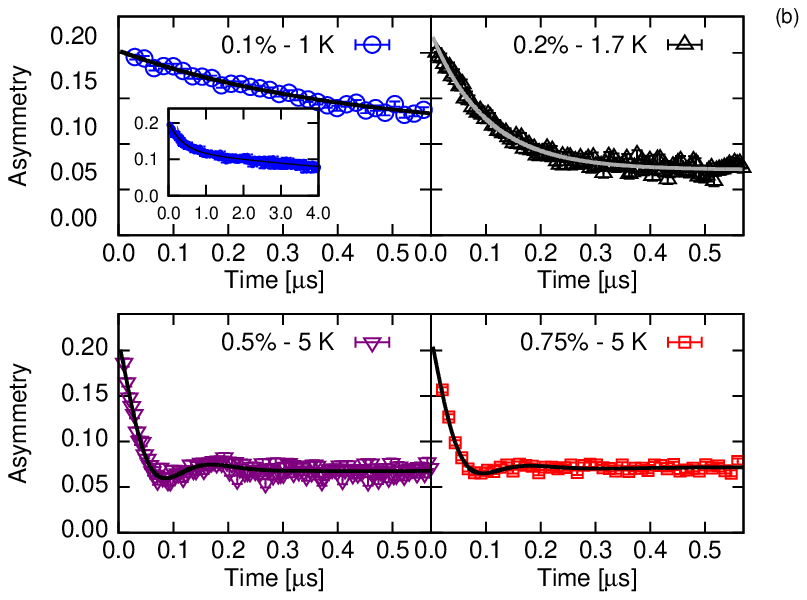} 
    \caption{(a) Asymmetry oscillations in the superconducting state of the samples with $x=0$\,\%, 0.025\,\%, and 0.075\,\% at low temperatures, measured in a transverse field of 200\,G. (b) ZF $\mu$SR time spectra at short time scales for $x=0.1$\,\%, 0.2\,\%, 0.5\,\%, and 0.75\,\%, measured at low temperatures. The inset in the upper left panel ($x=$0.1\,\%) shows the same data at a longer time scale. The specific temperatures are reported in the respective labels. Lines in (a) and (b) represent fits according to the functions described in the text.}\label{fig:asyms}
\end{figure*}

Up to $0.1$\,\% Mn content, \slrrt\ was measured at both peaks of
the \as-NQR spectrum. No apparent difference between the specific
relaxations of the two peaks could be observed, confirming that
the two local charge environments expressed in the double-peaked
NQR spectra indeed coexist at the nanoscale.\cite{LangPRL2010} 

At high temperatures \slrrt\ of the Mn undoped sample is nearly
flat. Upon increasing the Mn content, the relaxation becomes
slightly faster and begins to show an upturn towards lower
temperatures. Interestingly, the transition into the
superconducting state is only visible in the \slrrt\ data of the
samples with $0$\,\% and $0.025$\,\% Mn content, where \slrrt\
decreases below $T_c$, as expected. For the samples with
$0.075$\,\% and $0.1$\,\% Mn content, no signature of the
superconducting transition can be observed in \slrrt, which
displays an even steeper increase below $T_c$. Below 15\,K, also
the spin-lattice relaxation rate of the sample with the lowest Mn
content increases after the initial decrease below $T_c=25$\,K.

This enhancement of \slrrt\ upon Mn substitution is a signature of
growing magnetic fluctuations which are governing the relaxation
processes, even in the superconducting state. These magnetic
fluctuations seem to be uncorrelated with the pairing mechanism,
since they increase upon Mn substitution while $T_c$ is strongly
suppressed. The nature of these spin dynamics will be discussed in
detail in Section IV.

\subsection{Muon spin rotation and relaxation spectroscopy}

Muon spin rotation and relaxation spectroscopy ($\mu$SR) is one of
the most powerful techniques available to date to access the
magnetic properties in the presence of impurities, owing both to
the extreme sensitivity of the muon to small magnetic fields, to
its $I= {1}/{2}$ spin which simplifies the interaction scheme and
to the rapidly decaying nature of the magnetic dipolar interaction
with localized moments.

To investigate the low temperature electronic properties of the
sample series, we performed zero field (ZF), transverse field (TF)
and longitudinal field (LF) $\mu$SR at the Paul Sherrer Institut
(PSI) - Villigen (CH) with the GPS instrument of the $\pi$M3 beam
line. For these measurements, pressed pellets of the powdered
samples were prepared and mounted onto the sample holder using
mylar tape. Fig.~\ref{fig:asyms} displays a few representative
time domain spectra of the muon asymmetry, namely the time
evolution of the muon spin polarization, in the superconducting
state for TF measurements [Fig.~\ref{fig:asyms}(a)] and in ZF
measurements for the samples with static magnetism
[Fig.~\ref{fig:asyms}(b)].

In the magnetic phase, the ZF $\mu$SR asymmetry of powder samples can be written as:
\begin{equation}
    A(t) = \sum _{i} ^{N} A_{\perp}^{(i)} f^{(i)}(t, B^{(i)}) + A_{\parallel} e^{-\lambda_{\parallel } t} \, , \label{eq:asym}
\end{equation}
where $A^{(i)}_{\perp}$ and $A_{\parallel}$ represent the initial
amplitudes of the muon spin component perpendicular (transverse)
and parallel (longitudinal) to the local magnetic field $B^{(i)}$,
respectively. $f^{(i)}(t, B)$ describes the time dependence of the
transverse component and $\lambda_{\parallel }$ is the decay rate
of the longitudinal one. The index $i$ accounts for inequivalent
muon sites which usually are resolved only in the transverse
component.

\begin{figure}
\begin{center}
    \includegraphics[width=\columnwidth, clip]{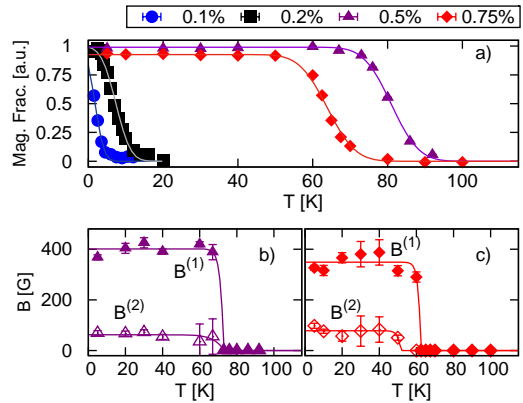} 
    \caption{(color online). (a) Magnetic volume fraction estimated from the longitudinal component of the ZF asymmetries (see text). 
    (b) and (c) Temperature dependence of the local magnetic fields $B^{(1,2)}_{\mu^{+}}$ (filled and empty symbols, respectively) at the muon site, detected by ZF $\mu$SR for $x = 0.5\%$ and 0.75\%, respectively. The lines are guides to the eye.}\label{fig:magprop}
\end{center}
\end{figure}

In a ZF experiment the field at the muon site $B^{(i)}$ can
originate only from the presence of spontaneous internal fields.
Let us first consider the ZF asymmetry for larger $x$ values:
The low temperature ZF $\mu$SR time signal of $x = 0.5$\,\% and
0.75\,\% [see Fig.~\ref{fig:asyms}(b)] show strongly damped
oscillations which reflect the muon spin precession around a
rather disordered distribution of local fields. The best fit
requires  $N=2$, in agreement with previous results on 1111,
\cite{Maeter2009,DeRenzi2012} with
$f_{\rm{ZF}}^{(i)}(t,B^{(i)})=\cos(2\pi \gamma B^{(i)} t)
e^{-(\lambda^{(i)}_{\perp} t)}$, and yields $\lambda^{(1)}_{\perp}\approx 20\,\mu s^{-1}$ and
$\lambda^{(2)}_{\perp}\approx 5\,\mu s^{-1}$. The temperature
evolution of the two local fields $B^{(i)}$ is displayed in
Figs.~\ref{fig:magprop}(b) and \ref{fig:magprop}(c).

Now we turn to the ZF asymmetry decay of the lower $x$ values.
The transverse component of the sample with $x$=0.2\,\% [Fig.~\ref{fig:asyms}(b)] displays only a fast decaying amplitude,
$f_{\rm{ZF}}(t)=e^{-(\lambda_{\perp} t)}$, with
$\lambda_{\perp}\approx 10\,\mu s^{-1}$. This is a signature of
overdamped oscillations due to the presence of a highly disordered
distribution of static internal fields with an amplitude $\Delta B =
\lambda_{\perp} / \pi \gamma \sim $ 200\,G (referring to the full
width at half maximum of the field distribution). The static
character of these fields is confirmed by LF measurements
performed at 1.5\,K (not shown), which reveal that an external
longitudinal field of the order of 1000\,G completely recovers the
muon spin polarization.

For the $x$=0.1\,\% sample [Fig.~\ref{fig:asyms}(b)] 
the transverse amplitude of the ZF time spectrum is sizeably reduced and displays an even
slower decay rate ($\lambda_{\perp} \sim 3\,\mu s^{-1}$), which
indicates a weakening of the magnetic state. For $x<0.1$\,\% no
transverse component is found but only an amplitude with a simple
gaussian decay rate due to nuclear dipolar interaction. For $x = 0.075$\,\%, this had to be multiplied by a tiny component with an exponential decay, arising from diluted magnetic impurities. This extra exponential decay rate was rouhgly constant.

For the magnetic samples $x\geq0.1$\,\% the magnetic volume fraction, i.e. the
fraction of the sample where muons detect a magnetic order, can be
evaluated as $V_{mag}= 3 (1-A_{\parallel}/A_{tot})/2
$,\cite{Sanna2013} with $A_{tot}$ being the total initial asymmetry
calibrated at high temperature. The temperature evolution of
$V_{mag}$ is displayed in Fig.~\ref{fig:magprop}(a) and shows that
a full magnetic volume fraction is achieved at low temperature for
$x\geq0.2$\,\%, while the $x=0.1$\,\% sample is only partially
magnetic. From these data it is possible to estimate the magnetic
transition temperature $T_N(x)$ (see Fig.~\ref{fig:phdiag}), which
can be empirically defined as the temperature at which
$V_{mag}=0.5$ .

It is noteworthy that for $x<0.1$\,\% no static magnetic state is
detected and the samples display only a superconducting character
below $T_c$. 

\begin{figure}[t]
    \includegraphics[width=0.7\columnwidth, clip]{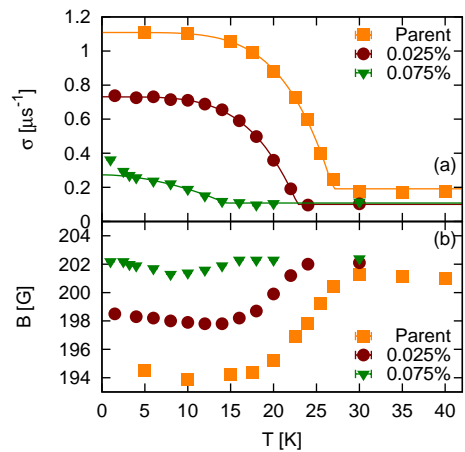} 
    \caption{(color online). (a) Muon spin relaxation rate $\sigma$ observed in transverse external fields in superconducting \lfmaso with $x=$0\,\%, 0.025\,\%, and 0.075\,\%. Lines are guides to the eyes. (b) Local field at the muon site for the same samples: the diamagnetic shift is clearly present in all superconducting compounds.}\label{fig:scprop}
\end{figure}

In order to further investigate the superconducting state, TF
$\mu$SR experiments have been performed by cooling the samples in
an external field of H=200\,G. In this case, since no spontaneous
random internal fields could be detected for $x<0.1$\,\%,
$A_{\parallel}^{(i)} = 0$ in Eq.~\eqref{eq:asym}. The fit of the
TF $\mu$SR signal [Fig.~\ref{fig:asyms}(a)] is described by
\begin{equation}
    f_{\rm{TF}}(t,B)= \cos(2\pi \gamma B t) e^{-(\sigma t)^2} \, ,
\end{equation}
where the Gaussian relaxation rate $\sigma$ below $T_c$ is
determined by the field distribution generated by the flux line
lattice.\cite{Sonier2000} Accordingly, in the clean limit,
$\sigma$ can be expressed in terms of the London penetration depth
$\lambda_L$, and turns out to be proportional to the supercarrier density
$n_{s}$:
\begin{equation}
    \sigma \propto \lambda_L ^{-2} \propto \frac{n_{s}}{m^{*}} \, , \label{eq:sigma}
\end{equation}
where $m^{*}$ is the effective mass of the
carriers.\cite{Brandt1988} The temperature evolution  both of
$\sigma(T)$ and $B(T)$ are displayed in Fig.~\ref{fig:scprop}.
Below $T_c$ a clear increase of $\sigma$ and a concomitant
diamagnetic shift of the local field $B = \mu_{0}H(1+\chi)$ (with
$\chi < 0$), characteristic of the superconducting ground state,
are observed.

The measurements highlight a strong decrease of $T_c$ upon Mn
substitution, in agreement with SQUID and detuning measurements.
Moreover, a decrease of the absolute value of $\sigma$ is observed
with increasing Mn content. According to Eq.~\eqref{eq:sigma} this
points towards a change of the superconducting carrier
concentration or of the effective mass (see Section IV for
details). The low temperature upturn of  both, $\sigma$ and
$B$, for the $x=0.075\%$ sample is possibly related to the growing
magnetic correlations detected by the $^{75}$As \slrr.

\section{Discussion}

The analysis of magnetic volumes when dealing with magnetic
impurities is a non trivial task and one must be careful in
distinguishing the various contributions to the ZF $\mu$SR
asymmetry.
As already mentioned, Mn impurities give rise to a static magnetic
state and Mn atoms likely participate in a short-range magnetic
order involving at least the neighboring Fe atoms which they
polarize. Given the high sensitivity of $\mu$SR to local fields,
it is possible that the ``magnetic islands'' surrounding the Mn
produce dipolar fields at the muon sites also outside the island
volume. This would result in a $\mu$SR signal with 100\% magnetic
volume (since all the muons probe a local field) but where the Fe
atoms would only partially be involved. We have checked whether this is indeed the
case for the $x=$0.1\,\%, 0.2\,\%, 0.5\,\% and
0.75\,\% samples  by performing simulations for the dipolar field
at the muon sites (see Appendix B) and evaluating the
correspondent time decay of the $\mu$SR asymmetry. While the rough
approximations used to tackle the problem do not allow definitive
conclusions for the samples with $x$=0.1\,\% and 0.2\,\%, for $x >
0.2$\,\% the simulations suggest that static magnetism develops
throughout the whole Fe plane. This observation is also supported by
the $T_N$ values which approach the ones of the undoped F-free
La1111, and can be hardly justified by a glassy ordering of a few
per thousand of Mn moments.

Our experimental results provide a microscopic insight into the
origin of the suppression of the superconducting ground state
already reported in Ref.~\onlinecite{SatoJPSJ2010}. Apart from the
rapid suppression of $T_c$, also a drastic change of the overall
temperature dependence of the resistivity upon adding Mn
impurities was reported in Ref.~\onlinecite{SatoJPSJ2010}. Already
a very small amount of Mn induces a significant upturn of the
resistivity at low temperatures, indicating a progressive
localization of charges and a concomitant transition to an
insulating ground state.\cite{SatoJPSJ2010} This
metal-to-insulator transition (MIT) is not expected since for such
a small amount of impurities, far below the Anderson localization
limit,\cite{Anderson1958} pair-breaking is expected to quench superconductivity but
to leave the system in a metallic state. Hence, the MIT and the
appearance of static magnetism in \lfmaso\ indicate a non-standard
origin for the weakening of superconductivity, likely due to the
proximity to a QCP.
\begin{figure}[t]
    \includegraphics[width=\columnwidth, clip]{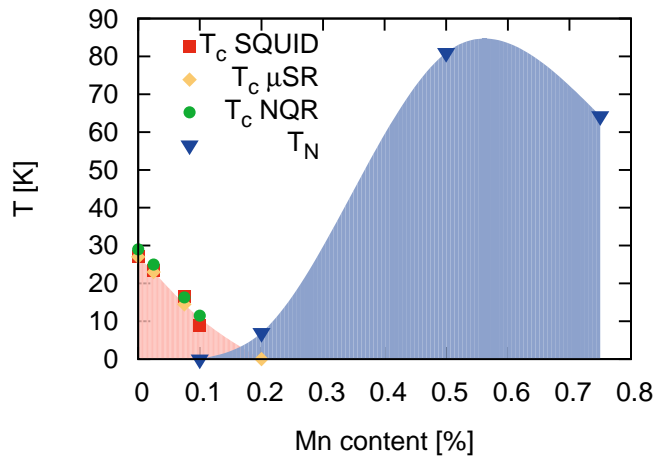} 
    \caption{(color online). Electronic phase diagram of \lfmaso. The dependence of the superconducting transition temperature $T_c$ on the magnetic impurities was determined from magnetization measurements (red squares),  $\mu$SR (orange triangles) and NQR (yellow diamonds). The magnetic transition temperature $T_N$ (blue triangles) was determined by $\mu$SR.}\label{fig:phdiag}
\end{figure}
This is also suggested by the electronic phase diagram which we
can extract from our $\mu$SR and NQR results (see
\F\ref{fig:phdiag}). Coherently with previous
reports\cite{SatoJPSJ2010} and magnetization measurements, $T_c$
is rapidly suppressed and superconductivity disappears for
x=0.2\,\%.
\begin{figure*}
\begin{center}
\includegraphics[width=\textwidth,clip]{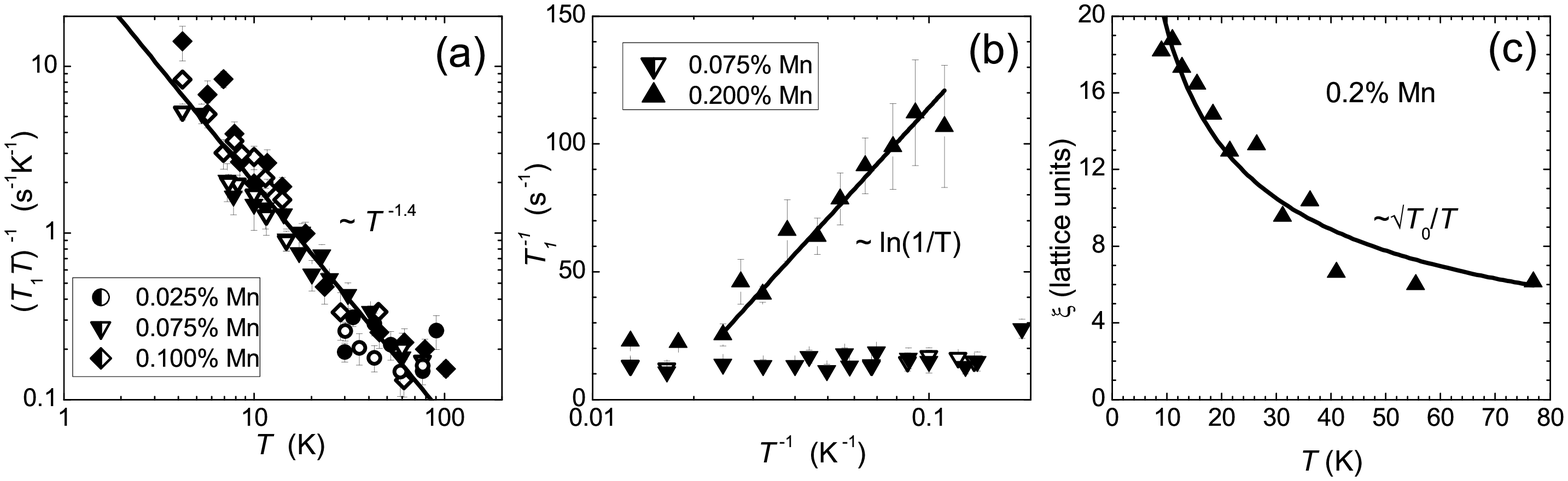} 
   \caption{(a) $(T_1T)^{-1}$ vs temperature for the samples with $0.025$\,\%, $0.075$\,\% and $0.1$\,\% Mn substitution. Filled and open symbols mark \slrrt\ measured on the low and high frequency peak of the double-peaked $^{75}$As-NQR spectrum, respectively. The solid line denotes the empirical power law dependence $(T_1T)^{-1} \propto T^{-1.4}$. (b)  \slrr\ vs inverse temperature for the samples with $0.075$\,\% (down-pointing triangles) and $0.2$\,\% (up-pointing triangles) Mn substitution, measured at the low (filled symbols) and high (open symbols) frequency peaks of the double-peaked $^{75}$As-NQR spectrum. The black solid line denotes the logarithmic temperature dependence $T_1^{-1} \propto \ln(1/T)$ for $x=0.2$\,\%, indicating 2D afm spin fluctuations. (c) Numerically calculated in plane correlation length of afm spin fluctuations for the same sample. The solid line shows $\xi \propto \sqrt{T_0/T}$.}\label{fig:T1analyzed}
\end{center}
\end{figure*}
Short range magnetism is observed in the $\mu$SR asymmetries for
$x$=0.1\,\% and 0.2\,\% (this hinders the observation of
superconductivity in the sample with $x=0.1$\,\% by means of
TF-$\mu$SR) while for $x > 0.2$\,\% the magnetic order develops through all the
FeAs plane. This order develops at the expenses of
superconductivity evidencing a strong competition between the two
ground states. Together with the charge localization probed by
resistivity measurements,\cite{SatoJPSJ2010} this points towards a
QCP at the boundary between the
superconducting and the magnetic ground state.

$^{75}$As NQR \slrr\ measurements evidence magnetic fluctuations for all
the superconducting Mn substituted samples which progressively
grow upon Mn substitution. To analyze the nature of the growing
spin fluctuations in this crossover region near the QCP, we
express the nuclear spin-lattice relaxation rate due to electronic
spin fluctuations as:\cite{Moriya1963, Ishigaki1996}
\begin{equation}
\frac{1}{T_1}=\frac{\gamma_n^2}{2} k_B T\frac{1}{N} \sum_{\vec{q}} |A_{\vec{q}}|^2\frac{\chi_{\perp}''(\vec{q},\omega_0)}{\omega_0} \, ,
\end{equation}
where $\gamma_n$ is the nuclear gyromagnetic ratio,
$A_{\vec{q}}$ the Fourier-$q$-component of the hyperfine coupling
constant, $\chi_{\perp}''(\vec{q},\omega_0)$ the imaginary part of
the dynamic susceptibility perpendicular to the quantization axis
of the nuclear spins (and thus perpendicular to the direction of
the EFG $z$ axes) and $\omega_0$ the nuclear Larmor frequency,
which can be basically taken as $\omega_0 \rightarrow 0$ since it
is much lower than the electron spin fluctuation frequency.

The measured \slrrt\ can be well described by a power law of the
form $(T_1T)^{-1} \propto T^{-b}$ with $b\simeq 1.4$, over a broad
doping and temperature range [see Fig.~\ref{fig:T1analyzed}(a)].
This is very alike to what has been observed in
\smfaof.\cite{Prando2010} In this compound, a similar increase of
$^{19}$F-NMR \slrrt\ has been observed due to a non-neglible
coupling between $f$ electrons and conduction electrons and has
been analyzed within the framework of the self-consistent
renormalization (SCR) theory. This justifies to analyze also our
data in the framework of the SCR theory, which is usually used to
describe spin fluctuations in weakly itinerant systems near a
QCP.

Based on the SCR theory, we calculated the spin-lattice
relaxation rate for both antiferromagnetic (afm) and ferromagnetic
(fm) spin fluctuations in two dimensions (see Appendix A).
Taking into account the resulting temperature dependencies of \slrr\ for
both cases [see Eqs.~\eqref{eq:T1afm} and \eqref{eq:T1fm}] it turns out
that the experimentally observed temperature dependence is determined by 2D antiferromagnetic spin fluctuations, which, next to a QCP, lead to \slrr$\propto \ln(1/T)$. In fact, this is indeed the  behavior observed for the sample with $x=0.2$\,\%, which is the closest to the QCP in the phase diagram [see
Fig.~\ref{fig:T1analyzed}(b)].

For this particular sample, the correlation length $\xi$
describing the in plane antiferromagnetic correlation can be
calculated. Starting from Eq.~\eqref{eq:T1afmcomplete} and
expressing the static susceptibility at the antiferromagnetic
wavevector $\chi(Q_{AF})$ in terms of the in plane correlation
length $\xi$:\cite{Prando2010}
\begin{equation}
\chi(Q_{AF}) = \frac{S(S+1)4\pi\xi^2}{3k_B T \ln(4\pi\xi^2+1)} \, ,
\end{equation}
the following dependence of the spin-lattice relaxation rate on the in plane correlation length results:
\begin{equation}
\frac{1}{T_1} = \frac{\gamma^2 A^2 \hbar S(S+1)}{4\pi 3k_B T_0} \frac{4\pi\xi^2}{\ln(4\pi\xi^2+1)} \, .
\label{eq:T1xi}
\end{equation}
By taking $A$=50\,kOe,\cite{GrafeNJP2009} and $S=1/2$, we have
derived $T_0\simeq 350$\,K from the high temperature limit, where
logarithmic corrections are not relevant and $\chi(Q_{AF})$
follows a simple Curie-Weiss behavior. The resulting numerically
calculated in plane correlation length for $x=0.2$\,\% is
plotted in Fig.~\ref{fig:T1analyzed}(c). Its temperature
dependence can be used for a double-check of the assumption of
antiferromagnetic spin fluctuations in the proximity to a QCP,
since in that case the in plane correlation
length should scale as $\xi \propto \sqrt{T_0/T}$ for $T \ll
T_0$.\cite{Prando2010, Millis1993} This is indeed what we find [see
Fig.~\ref{fig:T1analyzed}(c)] and confirms that the $^{75}$As-NQR
\slrrt\ is determined by 2D antiferromagnetic spin fluctuations.
Note that a recent $^{31}$P NMR study on LaFeAs$_{1-x}$P$_x$O also found evidence
for a quantum critical point in this compound expressed in strong antiferromagnetic fluctuations around $x$=0.3.\cite{Kitagawa2014} 

\begin{figure}[t]
    \includegraphics[width=0.9\columnwidth, clip]{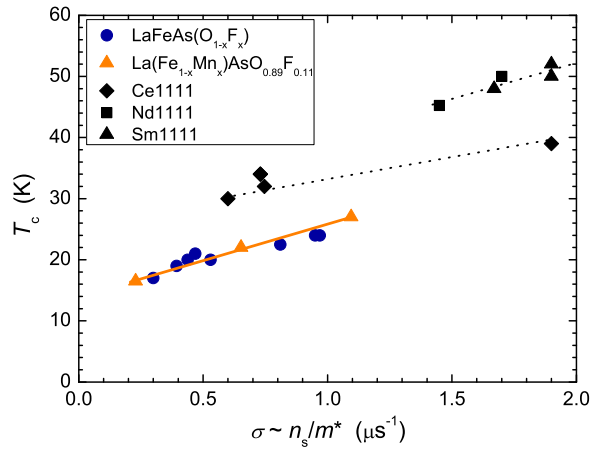} 
    \caption{(color online) Uemura plot, showing $T_c$ versus $\sigma \propto n_{\rm S}/m^{*}$ as deduced from our $\mu$SR results on \lfmaso\ (orange triangles), in comparison to \lafdoped\ (blue circles) and several other 1111 pnictide superconductors.\cite{Luetkens2008, Carlo2009, Carretta2013}}\label{fig:uplot}
\end{figure}

Further information on the effect of Mn in \laopt\ can be derived
by plotting the superconducting transition temperature $T_c$ vs
the TF-$\mu$SR Gaussian relaxation rate $\sigma \propto
n_{\rm S}/m^{*}$ [see Eq.~\eqref{eq:sigma}] which is usually known
as the Uemura plot.\cite{Uemura1989} Fig.~\ref{fig:uplot} shows
this plot for \lfmaso\ in comparison to several other 1111
iron-based superconductors.\cite{Luetkens2008, Carlo2009,
Carretta2013} Similarly to other compounds, a nice linear relation
between $T_c$ and $n_s/m^*$ is found also for \lfmaso. Remarkably,
$n_{\rm S}/m^{*}$ decreases even faster than $T_c(x)$,
most likely due to an enhancement of the effective mass $m^*$ upon
Mn substitution, since the system is approaching localization. Such an enhancement of $m^*$ has been recently reported for \bfamn\ and has been explained as a result of a Kondo-like band renormalization due to magnetic
scattering effects.\cite{Urata2013} Still, care should be taken when comparing the effects of impurities on 122 and 1111 iron-based superconductors, since they can differ a lot among different families of pnictides.
On the other hand, the similar behavior found in the Uemura plot
of \lfmaso\ and \lafdoped\ (Fig.~\ref{fig:uplot}) is likely to be a coincidence. For the latter, F-doping is known to cause an effective charge doping and thus should change the superconducting carrier density whereas, as
it has been shown by our $^{75}$As-NQR data, a change of the
carrier density by Mn for $x\leq 0.2$ \% is rather unlikely.
Direct measurements of the effective mass $m^*$ would help to
clarify this point.

Finally, it is worth noting that the extreme poisoning effect of
Mn is limited to La1111 only. For Nd 1111 and Sm 1111 the impurity
concentration leading to the suppression of superconductivity is
about ten times larger.\cite{SatoJPSJ2010, Singh2013} This
difference may originate from the details of the delicate {\it Ln}
dependence of the material parameters,
as highlighted in recent theoretical
works.\cite{Suzuki2013,Usui2012}

\section{Conclusion}

We studied the effect of tiny amounts of Mn impurities in \lfmaso,
which quenches superconductivity very effectively. Immediately
after the quench of $T_{c}$, static magnetism appears just
beside the superconducting dome. We showed that this magnetic
phase cannot involve just the diluted magnetic
impurities, but is intrinsic to the FeAs planes. Furthermore we
observed a progressive slowing down of spin fluctuations with
increasing Mn content, giving rise to an enhancement of \as\ NQR
\slrrt. The analysis of \slrrt\ showed that the spin fluctuations
are of 2D antiferromagnetic character and can be well described
within Moriya's SCR theory for weakly itinerant systems near a
quantum critical point. Together with the localization effects
found in resistivity measurements\cite{SatoJPSJ2010} we can
conclude that the effect of Mn impurities in \lfmaso\ goes beyond
a standard magnetic pair breaking effect and rather suggests the
proximity to a quantum critical point.

\section{Appendix}

\subsection{SCR theory of 2D spin fluctuations}

According to the SCR theory, the dynamical magnetic susceptibility in the paramagnetic phase in units of $(2\mu_B)^2$ is given by:\cite{Ishigaki1996}
\begin{equation}
\chi(q,\omega_0) = \frac{\pi T_0}{\alpha_Q T_A} \frac{x^\theta}{2\pi k_B T_0 x^\theta(y+x^2)-i \hbar \omega_0} \, ,
\label{eq:chi}
\end{equation}
with $T_0$ and $T_A$ being two parameters which characterize the width of the spin excitation spectrum in frequency and $q$ ranges, respectively, $\alpha_Q$ being a dimensionless interaction constant and:
\begin{equation}
y=\frac{1}{2\alpha_Q k_B T_A \chi(Q)} \, .
\label{eq:y}
\end{equation}
Furthermore, $x=\frac{q}{q_B}$, where $q_B$ is the effective zone boundary, and $\theta = 1$ and $0$ for ferromagnetic ($Q=0$) and antiferromagnetic ($Q \neq 0$) spin fluctuations, respectively.

\begin{figure*}
    \includegraphics[scale=0.5]{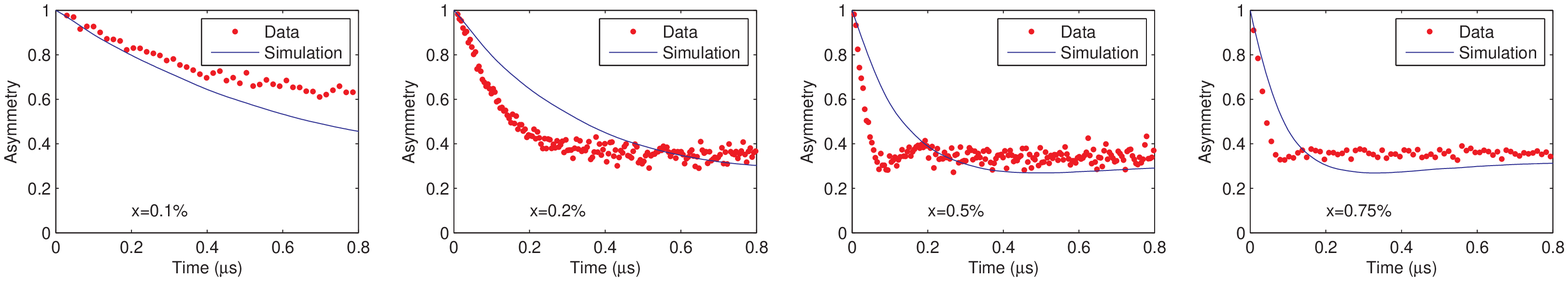} \\ 
    \includegraphics[scale=0.5]{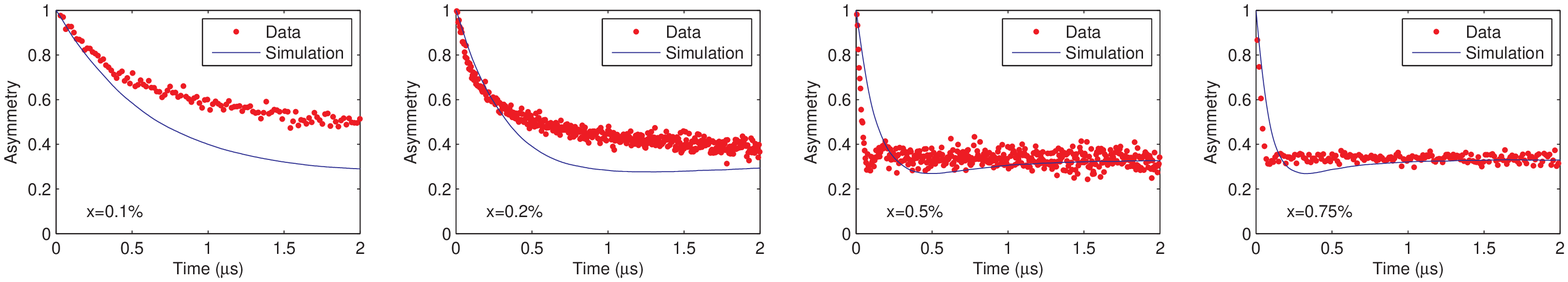} 
    \caption{(color online). Upper and lower panels show the experimental (red dots) and simulated (blue lines) asymmetry depolarizations for $x=0.1\,\% ,0.2\,\% ,0.5\,\%$ and 0.75\,\% on two different time scales. The experimental data are the same as in \F\ref{fig:asyms}(b).}\label{fig:calcasym}
\end{figure*}

We calculated the spin-lattice relaxation rate for both antiferromagnetic (afm) and ferromagnetic (fm) spin fluctuations in two dimensions, where $q_B = (4\pi/A_c)^{1/2}$ with $A_c$ being the unit cell volume.  $\chi_{\perp}''(\vec{q},\omega_0)/\omega_0$ was determined from Eq.~\eqref{eq:chi}. Assuming a $\vec{q}$-independent form factor $|A_{\vec{q}}|^2 = A^2$, as expected for itinerant systems, considering the limit $\omega_0 \rightarrow 0$ and integrating $\chi_{\perp}''(\vec{q},\omega_0)/\omega_0$ in two dimensions over a circle of radius $q_B$ one arrives at:
\begin{equation}
\frac{1}{T_1T} = \frac{\gamma_n^2 A^2}{2}\frac{\hbar}{4 \pi k_B}\frac{1}{\alpha_Q T_0 T_A}\frac{1}{y(y+1)}
\end{equation}
for antiferromagnetic fluctuations and
\begin{equation}
\frac{1}{T_1T} = \frac{\gamma_n^2 A^2}{2} \frac{\hbar}{4 \pi k_B}\frac{1}{\alpha_Q T_0 T_A} \left(\frac{1}{y(y+1)} + \frac{\tan^{-1}(1/\sqrt{y})}{y^{3/2}}\right)
\end{equation}
for ferromagnetic fluctuations. With Eq.~\eqref{eq:y} and assuming $T \ll T_A$, which implies $y \rightarrow 0$, the spin-lattice relaxation rate finally becomes:
\begin{equation}
\frac{1}{T_1} \simeq \frac{\hbar \gamma_n^2 A^2}{4 \pi}\left(\frac{T}{T_0}\right) \chi(Q_{AF}) \, \propto T\chi(Q_{AF})
\label{eq:T1afmcomplete}
\end{equation}
for the antiferromagnetic case and
\begin{equation}
\frac{1}{T_1} \simeq \frac{\hbar \gamma_n^2 A^2}{8}\sqrt{2 \alpha_Q k_B T_A}\left(\frac{T}{T_0}\right)\chi(0,0)^{3/2} \, \propto T\chi(0,0)^{3/2}
\end{equation}
for the ferromagnetic case.
The temperature dependence of the nuclear spin-lattice relaxation rate in the case of two dimensional antiferromagnetic/ferromagnetic spin fluctuations thus depends on the temperature dependence of the $\vec{q}$-specific susceptibility, which has been previously derived to scale as:\cite{Moriya2003}
\begin{eqnarray}
\chi(Q_{AF}) &\propto \dfrac{\ln\left(\tfrac{1}{T}\right)}{T}       \qquad &\text{for 2D afm}\\
\chi(0,0) &\propto \dfrac{1}{T \ln\left(\tfrac{1}{T}\right) } \qquad &\text{for 2D fm} \,.
\end{eqnarray}

We finally end up with the following temperature dependencies of the nuclear spin-lattice relaxation rate due to two dimensional spin fluctuations:
\begin{eqnarray}
\frac{1}{T_1} &\propto \ln\left(\tfrac{1}{T}\right)       \qquad &\text{for 2D afm} \label{eq:T1afm}\\
\frac{1}{T_1} &\propto \dfrac{1}{\sqrt{T}\left[\ln\left(\tfrac{1}{T}\right)\right]^{3/2}} \qquad &\text{for 2D fm} \label{eq:T1fm} \,.
\end{eqnarray}
The observed temperature dependence of the measured nuclear spin-lattice relaxation rate \slrr\, which is plotted in Fig.~\ref{fig:T1analyzed}(a), is clearly determined by 2D antiferromagnetic spin fluctuations. \slrr\ increases with decreasing temperature in the interesting temperature range, as suggested by Eq.~\eqref{eq:T1afm} and in particular, the data of the sample with $0.2$\,\% can be well fitted with a temperature dependence according to Eq.~\eqref{eq:T1afm}.\\

\subsection{Effect of diluted magnetic impurities}

To characterize the evolution of the magnetic ground state of \lfmaso 
as a result of Mn impurity substitution, we need to identify the contribution to the $\mu$SR signal due to the magnetic moment localized on the Mn and on the neighboring Fe atoms.
Indeed, given the high sensitivity of the $\mu$SR technique to small magnetic fields, strong magnetic moments diluted in the sample could give rise to a large volume fraction of muons probing a local field as a consequence of the dipolar interaction.

The presence of local moments on Mn atoms suggests that the impurities are surrounded by a small neighborhood of magnetic iron atoms, characterized by a short-ranged order. Nonetheless we do not have access to the magnetic moments on both Mn and Fe.
We are thus forced to a rough approximation to evaluate the $\mu$SR signal. We considered a large local moment of 3\,$\mu_{B}$ localized at Mn atoms' positions only. Even if this picture is un-physical since the Fe atoms do not participate to the static Mn order, it is a convenient and operative approximation to discriminate between the contributions coming from the magnetic states surrounding the impurities and those from the rest of the sample.

To estimate the field at the $\mu^{+}$ site we randomly substituted Mn impurities for Fe in the LaFeAsO structure with random local moment orientation. Only the dipolar interaction between the muon and the Mn impurities is considered.

The expected depolarization rates as a function of Mn
concentration are shown in Fig.~\ref{fig:calcasym}. As expected,
in the low dilution limit, the magnetic impurities give rise to an
exponential depolarization rate. For $x \geq 0.2\,\%$ a Lorentzian
Kubo-Toyabe-like trend is recovered.

For $x$=0.1\,\% the expected depolarization rate is rather close to the experimental values for $t<0.5\,\mu$s, but we note that a second slowly decaying component is present in the experimental signal. 
For $x=0.2\,\%$, the discrepancy of a factor of 2 between the data and the simulation does not allow conclusive inferences about the origin of the field at the muon sites. Nonetheless, for $x \geq 0.5\,\%$ the experimental depolarization rates are 3 to 6 times larger than the computationally estimated ones. Magnetic volumes surrounding the impurities are therefore much bigger than those in the $x=$0.1\,\% and 0.2\,\% samples and the presence of precessions strongly suggests that the whole iron plane is involved in the static magnetic ground state.

\section{Acknowledgments}
We thank M.~Mazzani, G.~Allodi and G.~Lang for discussion.
We acknowledge financial support from PSI EU funding. 
F.H., S.S., and P.C. acknowledge support by Fondazione Cariplo (Research Grant No. 2011-0266).


%

\end{document}